\documentclass[12pt]{article}

\usepackage{amssymb}
\usepackage{amsmath}
\usepackage[version=4]{mhchem}
\usepackage{xcolor}
\usepackage{graphicx}
\usepackage{a4wide}
\usepackage[square,numbers,compress]{natbib}
\usepackage{authblk}

\title{First-principles theory of electrochemical capacitance}

\author[1,*]{\small Tobias Binninger}
\affil[1]{ICGM, Univ Montpellier, CNRS, ENSCM, Montpellier,
France}
\affil[*]{Email: tobias.binninger.science@gmx.de}

\date{}

\begin{document}

\maketitle

\begin{abstract}
The differential capacitance comprises the most relevant thermodynamic information about an electrochemical system. Classical approaches to describe electrochemical capacitance have difficulties to combine the treatment of the ionic contribution of the electrolyte with the electronic contribution of the electrode. Moreover, different approaches are typically required for the description of the double-layer capacitance, on the one hand, and the pseudocapacitive contribution due to adsorption or intercalation of reactive species, on the other. In the present work, a new approach to describe electrochemical capacitance from first principles is developed. The treatment of a general electrochemical system at the level of multicomponent density-functional theory (MCDFT) yields an exact analytical expression for the total capacitance of the system, which corresponds to a formal series-circuit partitioning into ``quantum'' capacitance contributions of the density of states of electrons and ions, the (mean-field) electrostatic capacitance, as well as capacitance contributions of exchange and correlation among all active species. It is shown that the classical expression of the double-layer capacitance involving the interfacial Galvani potential is obtained in the limit of extended electrode and electrolyte regions. Importantly, the present formalism also describes systems with confined electrode and electrolyte phases, where the definition of the (classical) inner potentials of the electrode and electrolyte domains becomes problematic. Moreover, the new approach unifies the treatment of double-layer capacitance and pseudocapacitance resulting from reactive processes.
\end{abstract}

\section{Introduction}

The differential capacitance $C = \partial Q/\partial \mathcal{E}$ of an electrochemical system quantifies the linear response of the electrochemical charge $Q$ with respect to changes in electrode potential $\mathcal{E}$. The capacitance is also equal to the negative curvature of the grand potential (grand-canonical free energy) as a function of electrode potential~\cite{2010_book_Schmickler, 2019_JChemPhys_Hoermann_Marzari, 2022_JChemTheoComp_Binninger, 2022_EnergyEnvironSci_Binninger}, for which reason both the complete charging characteristics and the free energy curve of the system can be obtained by integration of the capacitance. Any electrochemical process leaves its trace in the capacitance, so the measurement of the latter, e.g. by cyclic voltammetry, is a central approach in experimental electrochemistry. Despite its central role, many questions around electrochemical capacitance remain to date. Recent experiments on platinum electrodes in highly diluted electrolyte solutions have revealed a significantly larger capacitance around the potential of zero charge (PZC) than expected from standard models~\cite{2022_PNAS_Koper}, thus raising new questions regarding the basic understanding of electrochemical double-layer capacitance. Renewed interest in the modeling of electrochemical interface capacitance further originates from the development of hybrid methods for the simulation of electrode--electrolyte interfaces, where a density-functional theory (DFT) description of the electrode material is coupled with classical continuum models for the electrolyte side~\cite{2006_PhysRevB_Otani, 2008_PhysRevB_Jinnouchi, 2010_arXiv_Dabo, 2012_PhysRevB_Letchworth-Weaver_Arias, 2017_PhysRevB_Otani, 
2017_JChemPhys_Sundararaman_Arias,
2018_JChemPhys_Schwarz,
2019_JChemPhys_Mathew_Hennig,
2019_JChemPhys_Nattino_Marzari,
2019_JChemPhys_Melander_Honkala, 2019_JChemPhys_Hoermann_Marzari, 2021_JPhysCondMat_Tesch_Kowalski_Eikerling, 2022_JChemTheoComp_Binninger}. 

Classical approaches towards modelling the capacitance of electrode--electrolyte interfaces rely on the definition of the inner electrostatic potentials $\phi_{\mathrm{etrode}}$ and $\phi_{\mathrm{elyte}}$ of the electrode and electrolyte, respectively~\cite{2010_book_Schmickler}. Changes in the electrode potential are identified with changes in the interfacial Galvani potential $\Delta\phi = \phi_{\mathrm{etrode}} - \phi_{\mathrm{elyte}}$ and the capacitance is the derivative of interfacial charge with respect to the latter, $C = \partial Q/\partial \Delta\phi$. The Gouy-Chapman-Stern (GCS) model~\cite{1924_ZElektrochem_Stern} underlies most of the current capacitance models, where the electrolytic part of the electrochemical double-layer is split into two regions: Fist, the inner Stern layer, which is bounded by the plane of closest approach of solvated ions (outer Helmholtz plane) giving rise to the Helmholtz capacitance $C_{\mathrm{H}}$. Dielectric saturation of the solvent at the interface, as well as the response of the electron density distribution to the excess charge at the electrode surface contribute to the Helmholtz capacitance~\cite{1983_JElectroanalChem_Badiali, 1989_Kornyshev_ElectrochimActa, 1996_ChemRev_Schmickler}. Second, the outer, or diffuse, layer giving rise to the Gouy-Chapman capacitance $C_{\mathrm{GC}}$. The series connection of these two capacitors yields the total double-layer capacitance $1/C_{\mathrm{dl}} = 1/C_{\mathrm{H}} + 1/C_{\mathrm{GC}}$. The GCS model predicts a capacitance minimum, caused by $C_{\mathrm{GC}}$, at the PZC and a convergence to $C_{\mathrm{H}}$ for increasing double-layer charge. Bikerman~\cite{1942_PhilosMag_Bikerman} included the effect of volume exclusion due to finite ion size, which results in a capacitance decrease at increased charging as a result of ion crowding~\cite{1997_PhysRevLett_Borukhov, 2007_JPhysChemB_Kornyshev}. Additional ion correlation terms that describe overscreening in the double-layer of ionic liquids have been introduced in the model by Bazant et al.~\cite{2011_PhysRevLett_Bazant_Kornyshev}. 

For electrodes with a strongly confined geometry, most notably one- or two-dimensional electrode materials, an additional capacitance contribution has been identified~\cite{1985_JPhysChem_Gerischer, 1988_ApplPhysLett_Luryi, 2004_JApplPhys_Pulfrey}, which is generally termed the quantum capacitance $C_{\mathrm{Q}}$. It is determined by the electronic density of states at the Fermi energy and responsible, e.g., for the characteristic v-shape of the capacitance vs. potential curve of graphene electrodes~\cite{2011_EnergyEnvSci_Stoller}. The question how to combine the description of quantum capacitance with the classical double-layer capacitance is relevant~\cite{2015_PhysRevB_Radin_Wood}. Both contributions are usually assumed to act in series, so the total capacitance is written $1/C = 1/C_{\mathrm{Q}} + 1/C_{\mathrm{dl}}$. I have recently shown~\cite{2021_PhysRevB_Binninger} that this relation can be rigorously derived from the formalism of joint density-functional theory (JDFT)~\cite{2005_JPhysChemB_Petrosyan_Arias, 2012_PhysRevB_Letchworth-Weaver_Arias}. Here, the electrode is treated at the (quantum-mechanical) level of Kohn-Sham DFT, whereas the electrolyte is described in terms of a classical density-functional theory. The hybrid quantum--classical formalism of JDFT has been derived by Petrosyan et al.~\cite{2005_JPhysChemB_Petrosyan_Arias} from the more fundamental formalism of multicomponent density-functional theory (MCDFT)~\cite{1982_JChemPhys_Capitani, 1998_PhysRevB_Gidopoulos, 2001_PhysRevLett_Gross, 2008_PhysRevLett_Hammes-Schiffer}, which is an extension of the (electronic) Hohenberg-Kohn DFT~\cite{1964_PhysRev_Hohenberg_Kohn} to include the treatment of nuclei at the same quantum-mechanical level as the electrons. However, to the best of my knowledge, MCDFT has never been directly employed to describe electrochemical capacitance so far.

Besides double-layer charging, reactive electrochemical processes contribute to the differential capacitance. Electrochemical adsorption of active species at an electrode surface, as well as electrochemical absorption, or intercalation, in the bulk of electrode materials are responsible for the experimentally observable pseudocapacitance~\cite{1991_JElectrochemSoc_Conway}. Specific adsorption has been included in classical continuum models~\cite{2016_JPhysChemC_Eikerling, 2017_JElectrochemSoc_Prendergast} with adsorption parameters that were fitted to experimental or DFT-computational results. In contrast, first-principles modeling of adsorption and intercalation generally requires an atomistic description of the active species. As an interesting alternative, MCDFT could enable a continuum treatment of reactive electrochemical processes from first principles.

In this article, I will show that the MCDFT formalism enables a unified first-principles description of electrochemical capacitance, including double-layer capacitance, quantum capacitance, and pseudocapacitance. The theory holds for the differential capacitance at electrochemical equilibrium. Extending my previous approach~\cite{2021_PhysRevB_Binninger}, an exact first-principles expression for the total capacitance is derived that precisely reduces to the expected forms in the classical limits for the double-layer capacitance and pseudocapacitance. As a great advantage, the presented formalism enables the treatment of electrochemical systems with strongly confined electrode and electrolyte phases, since it relies on a fundamental definition of the electrode potential that does \emph{not} require the definition of the respective inner potentials.

\section{Theory}

In section~\ref{subsec_JDFT_capacitance}, I will summarize and discuss the essential aspects of my recently presented approach~\cite{2021_PhysRevB_Binninger} for the unified description of quantum and classical capacitance of electrochemical interfaces within joint density-functional theory (JDFT). This will serve as the basis for the subsequent generalization to multicomponent density-functional theory (MCDFT) in section~\ref{subsec_MCDFT_capacitance}. This will lead us to a general formalism to describe the capacitance of electrochemical systems from first principles. 

\subsection{Quantum and classical capacitance from joint density-functional theory}
\label{subsec_JDFT_capacitance}

Joint density-functional theory (JDFT) combines the (quantum) DFT for the electronic subsystem with a classical DFT for the electrolyte~\cite{2005_JPhysChemB_Petrosyan_Arias, 2012_PhysRevB_Letchworth-Weaver_Arias}. I recently used this framework to derive the series-circuit partitioning of the total interface capacitance into quantum capacitance $C_{\mathrm{Q}}$, electrostatic double-layer capacitance $C_{\mathrm{dl}}$, and an additional capacitance $C_{\mathrm{xc}}$ resulting from the electronic exchange-correlation (XC) interaction~\cite{2021_PhysRevB_Binninger}: 
\begin{eqnarray}
\label{eq_capacitance_partitioning}
\frac{1}{C} = \frac{1}{C_{\mathrm{Q}}} + \frac{1}{C_{\mathrm{dl}}} + \frac{1}{C_{\mathrm{xc}}}
\end{eqnarray}

The starting point of the derivation is the JDFT expression for the Helmholtz free energy functional of the system at a temperature $T$~\cite{2012_PhysRevB_Letchworth-Weaver_Arias, 2017_JChemPhys_Sundararaman_Arias}:
\begin{eqnarray}
\label{eq_JDFT_Helmholtz_free_energy}
& A\left[n,\{n_i\}, \rho_{\mathrm{diel}}, \phi\right] \ =\ T^{\mathrm{\,ni}}_{\mathrm{kin}}[n] \,-\, TS^{\mathrm{ni}}[n] \,+\, F_{\mathrm{xc}}[n] \nonumber\\[0.2cm]
& \qquad + A^0_{\mathrm{elyte}}\left[\{n_i\}, \rho_{\mathrm{diel}}\right] \,-\, \frac{\epsilon_0}{2}\int|\nabla\phi|^2\,\mathrm{d}\mathbf{r} \nonumber\\[0.2cm]
& \qquad + \int \phi\,\left(\rho_{\mathrm{ext}}-e\,n+\sum eZ_i\,n_i+\rho_{\mathrm{diel}}\right) \mathrm{d}\mathbf{r} 
\end{eqnarray}
Here, $n(\mathbf{r})$, $\{n_i(\mathbf{r})\}$, $\rho_{\mathrm{diel}}(\mathbf{r})$, and $\rho_{\mathrm{ext}}(\mathbf{r})$ are the electron density, (electrolyte) ion densities, (solvent) dielectric charge density, and external charge density, respectively, where the latter represents the atomic cores of the electrode material. According to the Mermin-Kohn-Sham approach for electronic DFT at finite temperatures~\cite{1965_PhysRev_Mermin, 1965_PhysRev_Kohn_Sham}, the electronic part of the free energy functional~\eqref{eq_JDFT_Helmholtz_free_energy} is split into the kinetic energy $T^{\mathrm{\,ni}}_{\mathrm{kin}}[n]$ and entropy $S^{\mathrm{ni}}[n]$ of the noninteracting fermionic system and the exchange-correlation functional $F_{\mathrm{xc}}[n]$. $A^0_{\mathrm{elyte}}\left[\{n_i\}, \rho_{\mathrm{diel}}\right]$ is the free energy functional of the electrolyte at the level of classical DFT. The electrostatic potential $\phi(\mathbf{r})$ appears in expression~\eqref{eq_JDFT_Helmholtz_free_energy} as an auxiliary field and the corresponding free-energy minimization condition is the Poisson equation $\nabla^2\phi=-(\rho_{\mathrm{ext}}+\rho_{\mathrm{e}}+\rho_{\mathrm{ion}}+\rho_{\mathrm{diel}})/\epsilon_0$ with the solution:
\begin{eqnarray}
\label{eq_electrostatic_potential_solution}
\phi(\mathbf{r}) = \frac{1}{4\pi\epsilon_0}\,\int \frac{\left[\rho_{\mathrm{ext}}+\rho_{\mathrm{e}}+\rho_{\mathrm{ion}}+\rho_{\mathrm{diel}}\right](\mathbf{r}')}{|\mathbf{r}-\mathbf{r}'|}\,\mathrm{d}\mathbf{r}'
\end{eqnarray} 
Here, $\rho_{\mathrm{e}} = -e\,n$ and $\rho_{\mathrm{ion}} = \sum eZ_i\,n_i$ for ion species with charges $eZ_i$.

According to the variational principle, the equilibrium density distributions are obtained by minimizing the free energy functional of the appropriate ensemble. For electrochemical interfaces, the grand-canonical ensemble is typically considered most suitable, where the electrode is connected to an electronic reservoir controlling the electron chemical potential $\mu$ and the electrolyte is connected to an ionic reservoir determining the ion chemical potentials $\{\mu_i\}$. We therefore perform the Legendre transformation of the Helmholtz free energy $A$ to the grand potential $\Omega\ =\ A - \mu\,N - \sum \mu_i\,N_i$, where $N = \int n\,\mathrm{d}\mathbf{r}$ and $N_i = \int n_i\,\mathrm{d}\mathbf{r}$ are the electron and ion numbers, respectively. The variational minimization of the grand potential functional must be done under the constraint of overall charge neutrality of the electrochemical interface, $Q_{\mathrm{tot}}=\int (\rho_{\mathrm{ext}}+\rho_{\mathrm{e}}+\rho_{\mathrm{ion}}+\rho_{\mathrm{diel}})\,\mathrm{d}\mathbf{r} = 0$. Using the Lagrange multiplier method, the functional to be minimized reads $\Omega - \lambda \int (\rho_{\mathrm{ext}}+\rho_{\mathrm{e}}+\rho_{\mathrm{ion}}+\rho_{\mathrm{diel}})\,\mathrm{d}\mathbf{r}$. Minimization with respect to electron density variations yields the condition
\begin{eqnarray}
\label{eq_electronic_minimization}
\mu\ = \ \frac{\delta T^{\mathrm{\,ni}}_{\mathrm{kin}}}{\delta n(\mathbf{r})} - T\frac{\delta S^{\mathrm{ni}}}{\delta n(\mathbf{r})} + \frac{\delta F_{\mathrm{xc}}}{\delta n(\mathbf{r})} - e\,(\phi(\mathbf{r})-\lambda)
\end{eqnarray}
which is solved by the Kohn-Sham equations~\cite{1965_PhysRev_Kohn_Sham}
\begin{eqnarray}
\label{eq_KS_Schroedinger}
\left(-\frac{\hbar^2}{2m}\nabla^2 - e\,(\phi(\mathbf{r})-\lambda) + \mu_{\mathrm{xc}}(\mathbf{r}) \right)\psi_{\nu}(\mathbf{r}) = \epsilon_{\nu}\,\psi_{\nu}(\mathbf{r})
\end{eqnarray}
and
\begin{eqnarray}
\label{eq_KS_electron_density}
n(\mathbf{r}) = \sum_{\nu} \frac{1}{1+\exp\left(\frac{\epsilon_{\nu}-\mu}{kT}\right)} \,|\psi_{\nu}(\mathbf{r})|^2
\end{eqnarray}
Here, the electron density is given by the sum over the normalized Kohn-Sham orbital densities weighted with the Fermi-Dirac occupation numbers, and the XC-potential in equation~\eqref{eq_KS_Schroedinger} is given by $\mu_{\mathrm{xc}}(\mathbf{r}) = \delta F_{\mathrm{xc}}/\delta n(\mathbf{r})$. 

The meaning of the Lagrange multiplier $\lambda$ of the charge-neutrality constraint is a subtle, but relevant point. By considering the free-energy minimization conditions for the ion densities, one can show~\cite{2021_PhysRevB_Binninger} that the value of $\lambda$ must be equal to the plateau value $\phi_{\mathrm{elyte}}$ of the electrostatic potential~\eqref{eq_electrostatic_potential_solution} in the bulk electrolyte, $\lambda = \phi_{\mathrm{elyte}}$. The \emph{shifted} electrostatic potential $\phi(\mathbf{r})-\lambda = \phi(\mathbf{r})-\phi_{\mathrm{elyte}}$ in the Hamiltonian of equation~\eqref{eq_KS_Schroedinger} thus turns to zero in the bulk of the electrolyte, and the Kohn-Sham eigenvalues $\epsilon_{\nu}$ are referenced to the electrostatic energy of a ``solvated'' (test) electron in the bulk electrolyte. The latter is a commonly used reference state for the definition of the electrode potential $\mathcal{E}$~\cite{2012_PhysRevB_Letchworth-Weaver_Arias}, which then is simply given by the electron chemical potential $\mu$ that appears in the Fermi-Dirac distribution:
\begin{eqnarray}
\label{eq_electrode_potential_JDFT}
-e\,\mathcal{E} = \mu
\end{eqnarray}
It is emphasized that this definition of the electrode potential is valid for any electrode material, including semiconductors. Even in the presence of a band gap, the chemical potential $\mu$ of the Fermi-Dirac distribution, and thus also the electrode potential, is well-defined and continuous at any finite temperature $T>0\,\mathrm{K}$. For $T=0\,\mathrm{K}$, the limiting value of $T\rightarrow 0\,\mathrm{K}$ is naturally used~\cite{2021_PhysRevB_Binninger}.

Before proceeding, a clarifying note is required on the meaning of partial derivatives with respect to the electron number $N$. As discussed above, the grand-canonical description is most suited for electrochemical systems. Here, the electron number is not an externally controlled variable, but it is determined by the electron chemical potential and the other state variables (ion chemical potentials, temperature, volume, ...), i.e. $N(\mu,\{\mu_i\}, T, V)$. For fixed $\{\mu_i\}$, $T$, $V$, the relationship between $N$ and $\mu$ can be inverted and any thermodynamic quantity of the grand-canonical ensemble thus also expressed as a function of $N$. Partial derivatives with respect to $N$ are understood accordingly.

We are now in a position to obtain an explicit expression for the interface capacitance, following Ref.~\cite{2021_PhysRevB_Binninger}. Since changes in electron number correspond to changes in interfacial charge, $-e\,\mathrm{d}N = \mathrm{d}Q$, the interface capacitance $C$ is given by:
\begin{eqnarray}
\label{eq_capacitance_JDFT}
\frac{1}{C} = \frac{\partial \mathcal{E}}{\partial Q} = \frac{1}{e^2}\,\frac{\partial \mu}{\partial N}
\end{eqnarray}
Integrating equation~\eqref{eq_KS_electron_density} over space yields the total electron number
\begin{eqnarray}
\label{eq_electron_number_Fermi}
N = \sum_{\nu} \frac{1}{1+\exp\left(\frac{\epsilon_{\nu}-\mu}{kT}\right)}
\end{eqnarray}
We take the partial derivative of both sides with respect to $N$, and resolve the resulting equation for $\partial \mu/\partial N$ to obtain:
\begin{eqnarray}
\label{eq_chem_pot_deriv_I}
\frac{\partial \mu}{\partial N} = \frac{1}{g^T_{\mathrm{DOS}}(\mu)}\left(1 + \sum_{\nu} p^T(\mu - \epsilon_{\nu})\frac{\partial \epsilon_{\nu}}{\partial N}\right) 
\end{eqnarray}
Here, $g^T_{\mathrm{DOS}}(\epsilon) = \sum_{\nu} p^T(\epsilon-\epsilon_{\nu})$ is the temperature-dependent density of states (DOS), which is given by a series of thermally-broadened normalized peaks $p^T(x) = (1/kT)\,\exp[x/kT]/\left(1+\exp[x/kT]\right)^2$ centered at each of the Kohn-Sham eigenvalues $\epsilon_{\nu}$. We note, in passing, that $g^T_{\mathrm{DOS}}(\epsilon)$ represents a better, since physically meaningful, choice for DOS plotting purposes, rather than the typically used Gaussian broadening of a discrete eigenvalue spectrum.

As a last step, the derivatives of the eigenvalues in equation~\eqref{eq_chem_pot_deriv_I} are expressed by applying the Hellmann-Feynman theorem to equation~\eqref{eq_KS_Schroedinger}:
\begin{eqnarray}
\label{eq_KS_eigenval_deriv}
\frac{\partial \epsilon_{\nu}}{\partial N} = \int |\psi_{\nu}(\mathbf{r})|^2 \left((-e)\frac{\partial \left[\phi(\mathbf{r})-\phi_{\mathrm{elyte}}\right]}{\partial N} + \frac{\partial \mu_{\mathrm{xc}}(\mathbf{r})}{\partial N}\right) \mathrm{d}\mathbf{r} 
\end{eqnarray}
Inserting into~\eqref{eq_chem_pot_deriv_I} yields:
\begin{eqnarray}
\frac{\partial \mu}{\partial N}\, =\, \frac{1}{g^T_{\mathrm{DOS}}(\mu)} + \int \frac{g^T_{\mathrm{LDOS}}(\mu,\mathbf{r})}{g^T_{\mathrm{DOS}}(\mu)} \left((-e)\frac{\partial \left[\phi(\mathbf{r})-\phi_{\mathrm{elyte}}\right]}{\partial N} + \frac{\partial \mu_{\mathrm{xc}}(\mathbf{r})}{\partial N}\right)\mathrm{d}\mathbf{r} 
\label{eq_chem_pot_deriv_II}
\end{eqnarray}
Here, $g^T_{\mathrm{LDOS}}(\epsilon,\mathbf{r}) = \sum_{\nu} p^T(\epsilon-\epsilon_{\nu})|\psi_{\nu}(\mathbf{r})|^2$ is the temperature-dependent \emph{local} density of states (LDOS), fulfilling $\int g^T_{\mathrm{LDOS}}(\epsilon,\mathbf{r}) \mathrm{d}\mathbf{r} = g^T_{\mathrm{DOS}}(\epsilon)$. From equations~\eqref{eq_capacitance_JDFT} and \eqref{eq_chem_pot_deriv_II}, we immediately obtain the series-circuit partitioning of the total interface capacitance according to equation~\eqref{eq_capacitance_partitioning},
\begin{eqnarray}
\label{eq_total_capacitance_JDFT}
\frac{1}{C} \,=\, \frac{1}{e^2}\,\frac{\partial \mu}{\partial N} \,=\, \frac{1}{C_{\mathrm{Q}}} + \frac{1}{C_{\mathrm{dl}}} + \frac{1}{C_{\mathrm{xc}}}
\end{eqnarray}
where the quantum capacitance $C_{\mathrm{Q}}$ is given by the electronic DOS at the Fermi energy
\begin{eqnarray}
\label{eq_quantum_capacitance_JDFT}
C_{\mathrm{Q}} = e^2\,g^T_{\mathrm{DOS}}(\mu)
\end{eqnarray}
and the electrostatic double-layer capacitance $C_{\mathrm{dl}}$ and exchange-correlation capacitance $C_{\mathrm{xc}}$ are given by the expressions
\begin{eqnarray}
\label{eq_dl_capacitance_JDFT}
\frac{1}{C_{\mathrm{dl}}} = \frac{1}{e^2}\,\int \Gamma^T_{\mathrm{LDOS}}(\mu,\mathbf{r})\, (-e)\frac{\partial \left[\phi(\mathbf{r})-\phi_{\mathrm{elyte}}\right]}{\partial N} \,\mathrm{d}\mathbf{r} 
\end{eqnarray}
and
\begin{eqnarray}
\label{eq_xc_capacitance_JDFT}
\frac{1}{C_{\mathrm{xc}}} = \frac{1}{e^2}\,\int \Gamma^T_{\mathrm{LDOS}}(\mu,\mathbf{r})\, \frac{\partial \mu_{\mathrm{xc}}(\mathbf{r})}{\partial N} \,\mathrm{d}\mathbf{r} 
\end{eqnarray}
respectively. Here, $\Gamma^T_{\mathrm{LDOS}}(\mu,\mathbf{r}) = g^T_{\mathrm{LDOS}}(\mu,\mathbf{r}) / g^T_{\mathrm{DOS}}(\mu)$ is the \emph{normalized} LDOS at the Fermi energy.

\begin{figure}[t]
\centering
\includegraphics[width=0.4\textwidth]{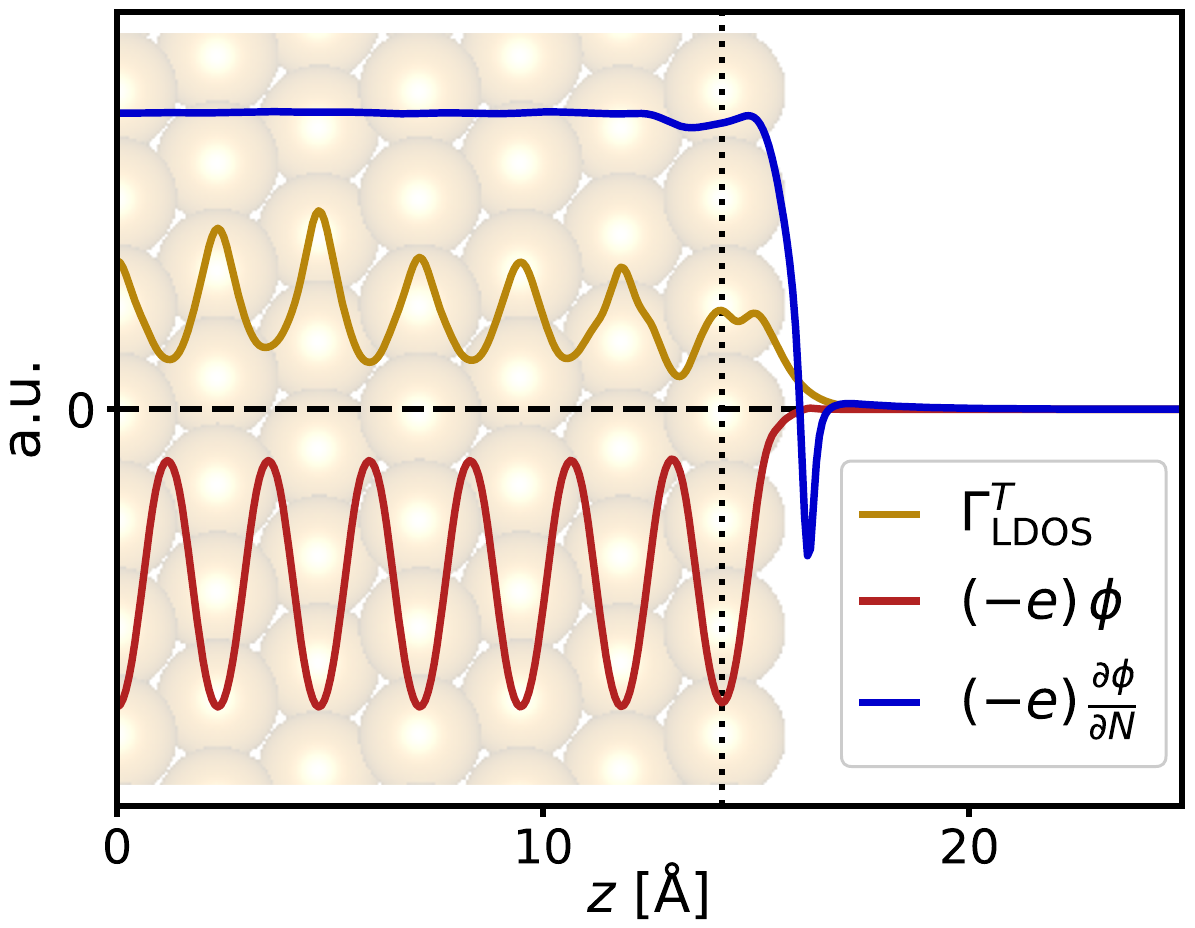}
\caption{Electrode bulk limit: Distribution of the (electronic) LDOS at the Fermi energy, $\Gamma^T_{\mathrm{LDOS}}(\mu,\mathbf{r})$, electrostatic potential $(-e)\,\phi(\mathbf{r})$, and potential derivative $(-e)\,\partial \phi(\mathbf{r}) / \partial N$ computed for a 13-layer Au (111) slab electrode. Plotted functions are in-plane averaged. Computational methods are described in the caption of Figure~\ref{fig_XC_capacitance}. Whereas the electrostatic potential $\phi(\mathbf{r})$ is strongly oscillating, its derivative $\partial \phi(\mathbf{r}) / \partial N$ assumes a constant plateau value in the bulk of the electrode.}
\label{fig_Au_bulk}
\end{figure}

\paragraph{Classical limit}
The analytically exact expressions of the individual capacitance contributions require some further discussion. Whereas expression~\eqref{eq_quantum_capacitance_JDFT} for the quantum capacitance precisely corresponds to the typically used definition~\cite{2007_ApplPhysLett_Fang, 2011_EnergyEnvSci_Stoller, 2015_PhysRevB_Radin_Wood, 2016_JPhysChemLett_Zhan, 2020_JElectroanalChem_Schmickler}, expression~\eqref{eq_dl_capacitance_JDFT} for the double-layer capacitance $C_{\mathrm{dl}}$ clearly differs from the classical form. In particular, the appearance of the local density of states in equation~\eqref{eq_dl_capacitance_JDFT} introduces additional ``quantum effects'' in $C_{\mathrm{dl}}$ that are not captured by the quantum capacitance $C_{\mathrm{Q}}$. In general, it would therefore be wrong to consider $C_{\mathrm{dl}}$ as a purely classical capacitance. However, in the limit of a bulk metal electrode, i.e. for large electrode thickness, it can be shown that the classical expression is recovered~\cite{2021_PhysRevB_Binninger}: Because $\Gamma^T_{\mathrm{LDOS}}$ is normalized, the integral in equation~\eqref{eq_dl_capacitance_JDFT} corresponds to a weighted average of the potential derivative $\partial \phi / \partial N$. Since the LDOS at the Fermi energy is nonzero in the bulk of an electronic conductor, the weight of the electrode bulk region in the integral of equation~\eqref{eq_dl_capacitance_JDFT} turns to one with increasing electrode thickness. This is confirmed by the computational results shown in Figure~\ref{fig_Au_bulk}, where the oscillating nature of the LDOS in a bulk gold electrode is apparent. Moreover, $\partial \phi(\mathbf{r}) / \partial N$, also shown in Figure~\ref{fig_Au_bulk}, reveals a constant plateau value $\partial \phi_{\mathrm{etrode}}/\partial N$ in the electrode bulk region. This immediately follows from the fact that the derivative $\partial \phi / \partial N$ fulfills the Poisson equation $\nabla^2(\partial \phi / \partial N)=-(\partial \rho / \partial N)/\epsilon_0$. Since, for an electrochemical interface, the charge density response $\partial \rho / \partial N$ is localized in the interfacial double-layer, the potential derivative $\partial \phi / \partial N$ becomes constant outside of the interfacial region. For a bulk metal electrode, we therefore have $\int \Gamma^T_{\mathrm{LDOS}}\left(\partial \phi/\partial N\right)\mathrm{d}\mathbf{r} \rightarrow \partial \phi_{\mathrm{etrode}}/\partial N$, and expression~\eqref{eq_dl_capacitance_JDFT} for $C_{\mathrm{dl}}$ turns to:
\begin{eqnarray}
\label{eq_dl_capacitance_JDFT_cl_limit}
\frac{1}{C_{\mathrm{dl}}} \,\rightarrow\, -\frac{1}{e}\,\frac{\partial \left[\phi_{\mathrm{etrode}}-\phi_{\mathrm{elyte}}\right]}{\partial N} = \frac{\partial \Delta\phi}{\partial Q}
\end{eqnarray} 
Here, we again used $-e\,\mathrm{d}N = \mathrm{d}Q$ and the fact that the normalized integral over the constant $\phi_{\mathrm{elyte}}$ simply reproduces that same value. $\Delta\phi = \phi_{\mathrm{etrode}}-\phi_{\mathrm{elyte}}$ is the difference between the inner potentials of the electrode and electrolyte, i.e. the Galvani potential of the electrochemical interface. The classical limit of the JDFT expression for $C_{\mathrm{dl}}$ thus yields the ``correct'' classical expression, where the inverse of the double-layer capacitance is given by the derivative of the interfacial Galvani potential $\Delta\phi$ with respect to the interfacial charge $Q$. We highlight that, in fact, the present formalism avoids the somewhat intricate definition of the inner potential $\phi_{\mathrm{etrode}}$, which typically requires performing some macroscopic average over the (nonconstant) microscopic potential $\phi(\mathbf{r})$ within the bulk of the electrode, see Figure~\ref{fig_Au_bulk}. Instead, here, we only require the existence of the bulk plateau value $\partial \phi_{\mathrm{etrode}}/\partial N$ of the potential \emph{derivative}, which is provided by the constancy of $\partial \phi(\mathbf{r}) / \partial N$ at a microscopic level in the bulk electrode region, as discussed above. 

So far, we have employed a quantum DFT description of the electrode, whilst treating the electrolyte in (simplified) classical terms. In particular, the existence of a constant plateau value $\phi_{\mathrm{elyte}}$ of the electrostatic potential in the bulk electrolyte region is required. This unsymmetric treatment of the electrode and electrolyte sides of the interface is somewhat unsatisfactory, which raises the question for the generalization of the formalism to achieve a symmetric first-principles description of the entire interface.

\subsection{General first-principles theory of electrochemical capacitance}
\label{subsec_MCDFT_capacitance}

Multicomponent density-functional theory (MCDFT)~\cite{1982_JChemPhys_Capitani, 1998_PhysRevB_Gidopoulos, 2001_PhysRevLett_Gross, 2008_PhysRevLett_Hammes-Schiffer} is a suitable theoretical framework for a symmetric first-principles description of electrochemical systems. It rests on the extension of the Hohenberg-Kohn theorem~\cite{1964_PhysRev_Hohenberg_Kohn} to the combined nonadiabatic quantum-mechanical treatment of electrons and nuclei (ions) introduced by Capitani, Nalewajski, and Parr~\cite{1982_JChemPhys_Capitani}, who proved the existence of the universal variational energy functional for a given multicomponent system. The corresponding extension of the Kohn-Sham approach has been presented by Gidopoulos~\cite{1998_PhysRevB_Gidopoulos}, who introduced the ion--ion exchange-correlation functional $F_{\mathrm{xc}}^{ii}[n_i]$ and the electron--ion correlation functional $F_{\mathrm{c}}^{ei}[n,n_i]$. 

\paragraph{MCDFT formalism for an electrochemical system}
We now apply the framework of MCDFT to an electrochemical system. The active electrochemical species are described at the MCDFT level. This comprises the electron density $n(\mathbf{r})$ and the ion densities $\{n_i(\mathbf{r})\}$. In contrast, the atomic cores of the electrode material, located at certain (fixed) positions, represent an external charge density that creates an external potential. In addition, certain less mobile ionic cores of the electrolyte can also be considered as part of the external charge density. The nonzero external potential also provides the required spatial anchoring for the MCDFT densities, which, otherwise, would be constant everywhere~\cite{1998_PhysRevB_Gidopoulos, 2001_PhysRevLett_Gross}. In principle, the solvent (nuclei) can be included at the MCDFT level. However, for the sake of clarity, we ignore these details in the following. 

The meaning of the ion densities $\{n_i(\mathbf{r})\}$ requires some clarification. At a basic first-principles level, the ion densities would correspond to the densities of the ionic nuclei, whereas the core electrons of the ions are part of the electron density $n(\mathbf{r})$, as in all-electron DFT. For the case of $\ce{H^+}$, there are no core electrons and $n_{\ce{H^+}}(\mathbf{r})$ simply corresponds to the density distribution of protons. For other ions such as $\ce{Li^+}$, it is desirable to combine the (first-principles) density of the respective nuclei, e.g. $n_{\ce{Li^{3+}}}$, with the corresponding core electron density to treat the system by the effective ion density, e.g. $n_{\ce{Li^+}}$. This very relevant technical aspect, however, is beyond the conceptual focus of the present work.

We combine the Kohn-Sham approach for MCDFT by Gidopoulos~\cite{1998_PhysRevB_Gidopoulos} with the generalization of (MC)DFT to thermodynamic ensembles by Mermin~\citep{1965_PhysRev_Mermin} to write the Helmholtz free energy functional of the system at a temperature $T$:
\begin{eqnarray}
\label{eq_MCDFT_Helmholtz_free_energy}
& A\left[n,\{n_i\}, \phi\right] \ =\ T^{\mathrm{\,ni}}_{\mathrm{kin}}[n] \,-\, TS^{\mathrm{ni}}[n] \,+\, F_{\mathrm{xc}}[n] \nonumber\\[0.2cm]
& \qquad + \sum_i \left\{ T^{\mathrm{\,ni}}_{\mathrm{kin},i}[n_i] \,-\, TS^{\mathrm{ni}}_i[n_i] \,+\, F_{\mathrm{xc}}^{ii}[n_i] \right\} + F_{\mathrm{c}}^{eI}[n,\{n_i\}] \nonumber\\[0.2cm]
& \qquad + \int \phi\,\left(\rho_{\mathrm{ext}}-e\,n+\sum_i eZ_i\,n_i\right) \mathrm{d}\mathbf{r} \,-\, \frac{\epsilon_0}{2}\int|\nabla\phi|^2\,\mathrm{d}\mathbf{r}
\end{eqnarray}
Here, for each of the ion species, the noninteracting kinetic energy and entropy functionals $T^{\mathrm{\,ni}}_{\mathrm{kin},i}[n_i]$ and $S^{\mathrm{ni}}_{i}[n_i]$, respectively, are introduced. The exchange-correlation energy contributions for each of the indistinguishable species are described by the functionals $F_{\mathrm{xc}}^{ii}[n_i]$. Finally, the correlation functional $F_{\mathrm{c}}^{eI}[n,\{n_i\}]$ captures all correlation effects among the different species. For the precise definition of these functionals, I refer to Gidopoulos~\cite{1998_PhysRevB_Gidopoulos}. As before, we introduce the electrostatic potential $\phi(\mathbf{r})$ as an auxiliary field and the respective variational minimization yields the Poisson equation with the solution~\eqref{eq_electrostatic_potential_MCDFT}. This approach is therefore precisely equivalent to explicitly introducing the electrostatic mean-field interactions between the charge densities in the free-energy functional. As before, we minimize the free energy in the grand-canonical ensemble for given chemical potentials $\mu$ and $\{\mu_i\}$ under the constraint of overall charge neutrality. The grand-potential functional to be minimized thus reads $A - \mu\,N - \sum \mu_i\,N_i - \lambda\,Q_{\mathrm{tot}}$, where $N = \int n\,\mathrm{d}\mathbf{r}$, $N_i = \int n_i\,\mathrm{d}\mathbf{r}$, $Q_{\mathrm{tot}}=\int (\rho_{\mathrm{ext}}-e\,n+\sum_i eZ_i\,n_i)\,\mathrm{d}\mathbf{r}$, and $\lambda$ is the Lagrange multiplier of the charge-neutrality constraint. Variational minimization leads to the conditions
\begin{eqnarray}
\label{eq_electronic_minimization_MCDFT}
\mu - e\,\lambda \ = \ \frac{\delta T^{\mathrm{\,ni}}_{\mathrm{kin}}}{\delta n(\mathbf{r})} - T\frac{\delta S^{\mathrm{ni}}}{\delta n(\mathbf{r})} + \frac{\delta F_{\mathrm{xc}}}{\delta n(\mathbf{r})} + \frac{\delta F_{\mathrm{c}}^{eI}}{\delta n(\mathbf{r})} - e\,\phi(\mathbf{r})
\end{eqnarray}
and
\begin{eqnarray}
\label{eq_ionic_minimization_MCDFT}
\mu_i + e Z_i\lambda\ = \ \frac{\delta T^{\mathrm{\,ni}}_{\mathrm{kin},i}}{\delta n_i(\mathbf{r})} - T\frac{\delta S^{\mathrm{ni}}_i}{\delta n_i(\mathbf{r})} + \frac{\delta F_{\mathrm{xc}}^{ii}}{\delta n_i(\mathbf{r})} + \frac{\delta F_{\mathrm{c}}^{eI}}{\delta n_i(\mathbf{r})} + e Z_i\,\phi(\mathbf{r})
\end{eqnarray}
which are solved by the Kohn-Sham equations~\cite{1965_PhysRev_Kohn_Sham, 1998_PhysRevB_Gidopoulos}
\begin{eqnarray}
\label{eq_KS_Schroedinger_MCDFT_electron}
\left(-\frac{\hbar^2}{2m}\nabla^2 - e\,\phi(\mathbf{r}) + \mu_{\mathrm{xc}}(\mathbf{r}) + \mu_{\mathrm{c},e}^{eI}(\mathbf{r}) \right)\psi_{\nu}(\mathbf{r}) = \epsilon_{\nu}\,\psi_{\nu}(\mathbf{r})
\end{eqnarray}
and
\begin{eqnarray}
\label{eq_KS_Schroedinger_MCDFT_ion}
\left(-\frac{\hbar^2}{2M_i}\nabla^2 + e Z_i\,\phi(\mathbf{r}) + \mu_{\mathrm{xc}}^{ii}(\mathbf{r}) + \mu_{\mathrm{c},i}^{eI}(\mathbf{r}) \right)\Psi^i_{\kappa}(\mathbf{r}) = \epsilon^i_{\kappa}\,\Psi^i_{\kappa}(\mathbf{r})
\end{eqnarray}
with the electrostatic potential
\begin{eqnarray}
\label{eq_electrostatic_potential_MCDFT}
\phi(\mathbf{r}) = \frac{1}{4\pi\epsilon_0}\,\int \frac{\left[\rho_{\mathrm{ext}}-e\,n+\sum_i eZ_i\,n_i\right](\mathbf{r}')}{|\mathbf{r}-\mathbf{r}'|}\,\mathrm{d}\mathbf{r}'
\end{eqnarray}
The correlation and exchange-correlation potentials in equations~\eqref{eq_KS_Schroedinger_MCDFT_electron} and \eqref{eq_KS_Schroedinger_MCDFT_ion} are defined as $\mu_{\mathrm{xc}}(\mathbf{r}) = \delta F_{\mathrm{xc}}/\delta n(\mathbf{r})$, $\mu_{\mathrm{xc}}^{ii}(\mathbf{r}) = \delta F_{\mathrm{xc}}^{ii}/\delta n_i(\mathbf{r})$, $\mu_{\mathrm{c},e}^{eI}(\mathbf{r}) = \delta F_{\mathrm{c}}^{eI}/\delta n(\mathbf{r})$, and $\mu_{\mathrm{c},i}^{eI}(\mathbf{r}) = \delta F_{\mathrm{c}}^{eI}/\delta n_i(\mathbf{r})$. The electron and ion densities are given by the weighted sum over the corresponding normalized Kohn-Sham orbital densities
\begin{eqnarray}
\label{eq_KS_electron_density_MCDFT}
n(\mathbf{r}) = \sum_{\nu} \frac{1}{1+\exp\left(\frac{\epsilon_{\nu}-\mu+e\,\lambda}{kT}\right)} \,|\psi_{\nu}(\mathbf{r})|^2
\end{eqnarray}
and
\begin{eqnarray}
\label{eq_KS_ion_density_MCDFT}
n_i(\mathbf{r}) = \sum_{\kappa} \frac{1}{1+\exp\left(\frac{\epsilon^i_{\kappa}-\mu_i-eZ_i\lambda}{kT}\right)} \,|\Psi^i_{\kappa}(\mathbf{r})|^2
\end{eqnarray}
Here, we assumed Fermi-Dirac statistics for the ion species, restricting the present treatment to fermionic ions. An extension to bosonic ions is foreseen in future works. 

Again, the Lagrange multiplier $\lambda$ plays a subtle role. In equations~\eqref{eq_electronic_minimization_MCDFT} and \eqref{eq_ionic_minimization_MCDFT}, we combined the respective terms with the chemical potentials on the left-hand side of the equations. Therefore, unlike in equation~\eqref{eq_KS_Schroedinger}, the Hamiltonians of equations~\eqref{eq_KS_Schroedinger_MCDFT_electron} and \eqref{eq_KS_Schroedinger_MCDFT_ion} do \emph{not} contain the potential shift $\lambda$, and the corresponding Kohn-Sham eigenvalue spectra are independent thereof. Instead, $\lambda$ now appears as a shift of the chemical potentials in the Fermi-Dirac distributions of equations~\eqref{eq_KS_electron_density_MCDFT} and \eqref{eq_KS_ion_density_MCDFT}. After integration over space, we obtain
\begin{eqnarray}
\label{eq_KS_electron_number_MCDFT}
N = \sum_{\nu} \frac{1}{1+\exp\left(\frac{\epsilon_{\nu}-\mu+e\,\lambda}{kT}\right)}
\end{eqnarray}
and
\begin{eqnarray}
\label{eq_KS_ion_number_MCDFT}
N_i = \sum_{\kappa} \frac{1}{1+\exp\left(\frac{\epsilon^i_{\kappa}-\mu_i-eZ_i\lambda}{kT}\right)}
\end{eqnarray}
We can now understand the importance of the gauge parameter $\lambda$. In the present grand-canonical formalism, the values of the chemical potentials $\mu$ and $\mu_i$ are externally controlled. In absence of the $\lambda$-shifts, the electron and ion numbers would be directly fixed by the values of the respective chemical potentials according to equations~\eqref{eq_KS_electron_number_MCDFT} and \eqref{eq_KS_ion_number_MCDFT}. In general, however, the resulting particle numbers would violate overall charge neutrality. The $\lambda$ terms in the Fermi-Dirac distributions provide the possibility to shift the Kohn-Sham eigenvalue spectra relative to the externally fixed chemical potentials. This enables tuning the electron and ion numbers in order to fulfill charge neutrality. We denote the suitably shifted chemical potentials as \emph{effective} chemical potentials, or effective Fermi energies, $\mu^{\mathrm{eff}} = \mu - e\,\lambda$ and $\mu_i^{\mathrm{eff}} = \mu_i + e Z_i\lambda$. 

\paragraph{Definition of the electrode potential}
To derive an expression for the capacitance, we first need to reconsider the definition of the electrode potential. In the JDFT framework, it is defined by equation~\eqref{eq_electrode_potential_JDFT}, or, more generally, by $-e\,\mathcal{E} = \mu - (-e)\phi_{\mathrm{elyte}}$, where the inner potential of the electrolyte serves as a reference~\cite{2012_PhysRevB_Letchworth-Weaver_Arias}. In the present formalism, however, we do \emph{not} require the existence of a plateau $\phi_{\mathrm{elyte}}$ of the electrostatic potential~\eqref{eq_electrostatic_potential_MCDFT} in the bulk electrolyte. Also, to preserve the first-principles character of our treatment, we wish to avoid any definition of the inner potential in terms of macroscopic averages. To this end, we resort to a real experimental situation, where the electrode potential is the measured voltage between working and reference electrode, $-e\,\mathcal{E} = \mu - \mu^{\mathrm{Ref}}$, i.e. the difference between the electron chemical potential $\mu$ of the working electrode and the electron chemical potential $\mu^{\mathrm{Ref}}$ of the reference electrode~\cite{2010_book_Schmickler}. The latter, in turn, is fixed by the electrochemical equilibrium of the given reference electrode reaction $\sum_i \alpha_i \mathrm{A}_i \rightleftharpoons \mathrm{e^-}$. Here, all reactant and product species $\mathrm{A}_i$ are written on the same side of the reaction equation and the corresponding stoichiometric coefficients are positive or negative for reduced or oxidized species, respectively. Because of charge conservation, the coefficients must fulfill 
\begin{eqnarray}
\label{eq_ref_charge_conservation}
\sum_i \alpha_i Z_i = -1
\end{eqnarray}
Electrochemical equilibrium at the reference electrode means $\mu^{\mathrm{Ref}} = \sum_i \alpha_i \mu_i$, so the (working) electrode potential is given by
\begin{eqnarray}
\label{eq_electrode_potential_MCDFT}
-e\,\mathcal{E} \,=\, \mu - \mu^{\mathrm{Ref}} \,=\, \mu - \sum_i \alpha_i \mu_i \,=\, \mu^{\mathrm{eff}} - \sum_i \alpha_i \mu_i^{\mathrm{eff}}
\end{eqnarray}
In the last step, we used the charge conservation condition for the stoichiometric coefficients to express the electrode potential in terms of the effective chemical potentials of the Fermi-Dirac distributions introduced above.

\paragraph{Capacitance}
The total capacitance of the electrochemical system is then given by:
\begin{eqnarray}
\frac{1}{C} \,=\, \frac{\partial \mathcal{E}}{\partial Q} \,=\, \frac{1}{e^2}\frac{\partial \mu^{\mathrm{eff}}}{\partial N} - \sum_i \alpha_i\, \frac{1}{e^2}\frac{\partial \mu_i^{\mathrm{eff}}}{\partial N} \nonumber \\[0.2cm]
= \frac{1}{e^2}\frac{\partial \mu^{\mathrm{eff}}}{\partial N} - \sum_i \alpha_i\,t_i\, \frac{1}{e^2}\frac{\partial \mu_i^{\mathrm{eff}}}{\partial N_i}
\label{eq_capacitance_MCDFT}
\end{eqnarray}
Here, as before, we defined changes in electrochemical charge by changes in electron number, $\mathrm{d}Q = -e\,\mathrm{d}N$. This is consistent with an experimental situation where changes in charge are quantified via the measured electronic current. In the last step, we introduced the coefficients $t_i = \left(\partial N_i/\partial N\right)_{\{\mu_i\}}$, which correspond to the response of the ion (particle) numbers to changes in electron number. There is an analogy between these coefficients and the ion transference numbers of electrolyte solutions, which specify the fraction of the electric current carried by each of the ion species. Because of overall charge neutrality, the coefficients $t_i$ must fulfill
\begin{eqnarray}
\label{eq_transference_coefficients_sum}
\sum_i Z_i\,t_i = 1
\end{eqnarray}

As before (cf. equations~\eqref{eq_chem_pot_deriv_I}, \eqref{eq_KS_eigenval_deriv}, \eqref{eq_chem_pot_deriv_II}), we obtain the particle-number derivatives in equation~\eqref{eq_capacitance_MCDFT} by first differentiating equations~\eqref{eq_KS_electron_number_MCDFT} and \eqref{eq_KS_ion_number_MCDFT} and then applying the Hellmann-Feynman theorem to the electronic and ionic Kohn-Sham equations~\eqref{eq_KS_Schroedinger_MCDFT_electron} and \eqref{eq_KS_Schroedinger_MCDFT_ion}:
\begin{eqnarray}
\frac{\partial \mu^{\mathrm{eff}}}{\partial N}\, =\, \frac{1}{g^T_{\mathrm{DOS}}(\mu^{\mathrm{eff}})} + \int \Gamma^T_{\mathrm{LDOS}}(\mu^{\mathrm{eff}},\mathbf{r})\left(-e\frac{\partial \phi(\mathbf{r})}{\partial N} + \frac{\partial \mu_{\mathrm{xc}}(\mathbf{r})}{\partial N} + \frac{\partial \mu_{\mathrm{c},e}^{eI}(\mathbf{r})}{\partial N}\right)\mathrm{d}\mathbf{r} 
\label{eq_chem_pot_deriv_electron_MCDFT}
\end{eqnarray}
and
\begin{eqnarray}
\frac{\partial \mu_i^{\mathrm{eff}}}{\partial N_i}\, =\, \frac{1}{g^T_{\mathrm{DOS},i}(\mu_i^{\mathrm{eff}})} + \int \Gamma^T_{\mathrm{LDOS},i}(\mu_i^{\mathrm{eff}},\mathbf{r})\left(e Z_i\frac{\partial \phi(\mathbf{r})}{\partial N_i} + \frac{\partial \mu_{\mathrm{xc}}^{ii}(\mathbf{r})}{\partial N_i} + \frac{\partial \mu_{\mathrm{c},i}^{eI}(\mathbf{r})}{\partial N_i}\right)\mathrm{d}\mathbf{r} 
\label{eq_chem_pot_deriv_ion_MCDFT}
\end{eqnarray}
Here, we introduced the (temperature-dependent) density of states $g^T_{\mathrm{DOS},i}(\epsilon) = \sum_{\kappa} p^T(\epsilon-\epsilon^i_{\kappa})$, local density of states $g^T_{\mathrm{LDOS},i}(\epsilon,\mathbf{r}) = \sum_{\kappa} p^T(\epsilon-\epsilon^i_{\kappa})|\Psi^i_{\kappa}(\mathbf{r})|^2$, and normalized LDOS $\Gamma^T_{\mathrm{LDOS},i}(\epsilon,\mathbf{r}) = g^T_{\mathrm{LDOS},i}(\epsilon,\mathbf{r}) / g^T_{\mathrm{DOS},i}(\epsilon)$ for each of the ion species. Inserting equations~\eqref{eq_chem_pot_deriv_electron_MCDFT} and \eqref{eq_chem_pot_deriv_ion_MCDFT} into \eqref{eq_capacitance_MCDFT}, we obtain the partitioning of the total capacitance:
\begin{eqnarray}
\label{eq_total_capacitance_MCDFT}
\frac{1}{C} \,=\, \frac{1}{C_{\mathrm{DOS}}} + \frac{1}{C_{\mathrm{el}}} + \frac{1}{C_{\mathrm{xc}}}
\end{eqnarray}
The total quantum, or DOS, capacitance $C_{\mathrm{DOS}}$ is now given by the contributions of all species' DOS at the respective Fermi energies:
\begin{eqnarray}
\label{eq_DOS_capacitance_MCDFT}
\frac{1}{C_{\mathrm{DOS}}} \,=\, \frac{1}{C_{\mathrm{DOS},e}} - \sum_i \frac{\alpha_i\,t_i}{C_{\mathrm{DOS},i}}
\end{eqnarray}
where $C_{\mathrm{DOS},e} = e^2\,g^T_{\mathrm{DOS}}(\mu^{\mathrm{eff}})$ and $C_{\mathrm{DOS},i} = e^2\,g^T_{\mathrm{DOS},i}(\mu_i^{\mathrm{eff}})$. The electronic DOS capacitance $C_{\mathrm{DOS},e}$ is the same as the (electronic) quantum capacitance $C_{\mathrm{Q}}$ introduced earlier. However, the ionic counterpart $C_{\mathrm{DOS},i}$ includes effects of the configurational density of states that are typically considered as classical, as discussed in section~\ref{subsec_discussion_intercalation}. Therefore, the more general term ``DOS capacitance'' is preferred. The electrostatic capacitance $C_{\mathrm{el}}$ reads
\begin{eqnarray}
\label{eq_el_capacitance_MCDFT}
\frac{-e}{C_{\mathrm{el}}} = \int \Gamma^T_{\mathrm{LDOS}}(\mu^{\mathrm{eff}},\mathbf{r})\frac{\partial \phi(\mathbf{r})}{\partial N}\mathrm{d}\mathbf{r} + \sum_i \alpha_i Z_i\int \Gamma^T_{\mathrm{LDOS},i}(\mu_i^{\mathrm{eff}},\mathbf{r}) \frac{\partial \phi(\mathbf{r})}{\partial N} \mathrm{d}\mathbf{r}
\end{eqnarray}
where we used $t_i\,\frac{\partial \phi(\mathbf{r})}{\partial N_i} = \frac{\partial N_i}{\partial N}\,\frac{\partial \phi(\mathbf{r})}{\partial N_i} = \frac{\partial \phi(\mathbf{r})}{\partial N}$. In principle, $C_{\mathrm{el}}$ corresponds to the electrostatic double-layer capacitance $C_{\mathrm{dl}}$ of equation~\eqref{eq_dl_capacitance_JDFT}. However, as discussed later, the present MCDFT formalism goes beyond the treatment of electrochemical interfaces, but also includes charging processes of electrochemical systems that do not have a double-layer structure, such as ion intercalation in active battery materials. The more general term ``electrostatic capacitance'' is therefore preferred. It must be emphasized that $C_{\mathrm{el}}$ only represents the capacitance contribution from mean-field electrostatics that are captured by the electrostatic potential $\phi(\mathbf{r})$. Effects due to electrostatic correlation and the correction of electrostatic self-interaction errors are part of the exchange-correlation capacitance $C_{\mathrm{xc}}$. The latter is given by:
\begin{eqnarray}
\label{eq_xc_capacitance_MCDFT}
\frac{1}{C_{\mathrm{xc}}} = \frac{1}{C_{\mathrm{xc}}^{ee}} + \frac{1}{C_{\mathrm{c}}^{eI}} - \sum_i \frac{\alpha_i\,t_i}{C_{\mathrm{xc}}^{ii}}
\end{eqnarray}
Here, 
\begin{eqnarray}
\label{eq_xc_ee_capacitance_MCDFT}
\frac{1}{C_{\mathrm{xc}}^{ee}} = \frac{1}{e^2}\int \Gamma^T_{\mathrm{LDOS}}(\mu^{\mathrm{eff}},\mathbf{r}) \,\frac{\partial \mu_{\mathrm{xc}}(\mathbf{r})}{\partial N}\,\mathrm{d}\mathbf{r}
\end{eqnarray}
and
\begin{eqnarray}
\label{eq_xc_ii_capacitance_MCDFT}
\frac{1}{C_{\mathrm{xc}}^{ii}} = \frac{1}{e^2}\int \Gamma^T_{\mathrm{LDOS},i}(\mu_i^{\mathrm{eff}},\mathbf{r})\,\frac{\partial \mu_{\mathrm{xc}}^{ii}(\mathbf{r})}{\partial N_i}\,\mathrm{d}\mathbf{r}
\end{eqnarray}
are the electron--electron and (like) ion--ion exchange-correlation capacitance contributions, and
\begin{eqnarray}
\label{eq_xc_ei_capacitance_MCDFT}
\frac{1}{C_{\mathrm{c}}^{eI}} = \frac{1}{e^2}\int \Gamma^T_{\mathrm{LDOS}}(\mu^{\mathrm{eff}},\mathbf{r})\,\frac{\partial \mu_{\mathrm{c},e}^{eI}(\mathbf{r})}{\partial N}\,\mathrm{d}\mathbf{r} - \sum_i \frac{\alpha_i\,t_i}{e^2}\int \Gamma^T_{\mathrm{LDOS},i}(\mu_i^{\mathrm{eff}},\mathbf{r})\,\frac{\partial \mu_{\mathrm{c},i}^{eI}(\mathbf{r})}{\partial N_i}\,\mathrm{d}\mathbf{r}
\end{eqnarray}
is the capacitance contribution due to correlation between the different species (electrons and ions). Equations~\eqref{eq_total_capacitance_MCDFT} to \eqref{eq_xc_ei_capacitance_MCDFT} are exact first-principles expressions for the individual contributions to the total capacitance of an electrochemical system, which is the central result of this article. They will be exemplified and discussed in more detail in the following section.

\section{Discussion}
\label{sec_discussion}

\subsection{Electrostatic double-layer capacitance and classical limit}
As in the JDFT expression~\eqref{eq_dl_capacitance_JDFT}, the electrostatic capacitance $C_{\mathrm{el}}$ of equation~\eqref{eq_el_capacitance_MCDFT} differs significantly from the classical picture. It comprises the average (integrals) of the electrostatic potential response $\partial \phi(\mathbf{r})/\partial N$ weighted with the normalized LDOS of electrons and each ion species. Remarkably, the capacitance, in general, depends on the choice of the reference electrode via the stoichiometric coefficients $\alpha_i$. We will see, however, that this dependence vanishes in the classical limit. In the JDFT framework, the classical capacitance expression was recovered in the limit of a bulk electrode, while the electrolyte was implicitly treated as a bulk reservoir. The present formalism, however, does not necessarily require the electrolyte to be an extended bulk phase. For the classical limit, we therefore need to consider the bulk limits of both the electrode and the electrolyte. Following the same arguments as before (see equation~\eqref{eq_dl_capacitance_JDFT_cl_limit}), the first integral in equation~\eqref{eq_el_capacitance_MCDFT} then turns to the constant plateau value $\partial \phi_{\mathrm{etrode}}/\partial N$ in the electrode. Similarly, since all the (normalized) LDOS $\Gamma^T_{\mathrm{LDOS},i}$ of the ion species are nonzero in the bulk of the electrolyte region, all corresponding integrals in equation~\eqref{eq_el_capacitance_MCDFT} turn to the same constant plateau value $\partial \phi_{\mathrm{elyte}}/\partial N$ in the electrolyte. As discussed earlier, $\partial \phi(\mathbf{r})/\partial N$ reaches the respective (constant) plateau values, because the charge density response $\partial \rho(\mathbf{r})/\partial N$ is localized around the interface and decays to zero in the bulk electrode and electrolyte regions. We therefore find:
\begin{eqnarray}
\label{eq_el_capacitance_MCDFT_cl_limit}
\frac{-e}{C_{\mathrm{el}}} \,\rightarrow\, \frac{\partial \phi_{\mathrm{etrode}}}{\partial N} + \sum_i \alpha_i Z_i\,\frac{\partial \phi_{\mathrm{elyte}}}{\partial N} = \frac{\partial \phi_{\mathrm{etrode}}}{\partial N} - \frac{\partial \phi_{\mathrm{elyte}}}{\partial N}
\end{eqnarray}
where we used the charge conservation condition~\eqref{eq_ref_charge_conservation} of the reference electrode reaction. We thus obtain the ``correct'' classical expression $1/C_{\mathrm{el}} = \partial \Delta\phi/\partial Q$ with the interfacial Galvani potential $\Delta\phi$. However, it should be emphasized that the present treatment does not rely on the definition of the (macroscopic) inner potentials of the electrode and electrolyte phases.

\subsection{DOS capacitance and confined electrolyte}

The electronic DOS capacitance, or quantum capacitance, is well-known for the case of confined electrode materials, such as single-layer graphene, with a limited electronic DOS at the Fermi level~\cite{2021_PhysRevB_Binninger}. In the same way, the ionic DOS capacitance $C_{\mathrm{DOS},i} = e^2\,g^T_{\mathrm{DOS},i}(\mu_i^{\mathrm{eff}})$ is given by the value of the ionic density of states at the corresponding ionic Fermi level. This analogy is nicely exemplified for the case of an aqueous electrolyte. It is well accepted that, from an electric perspective, water can be described as a protonic semiconductor~\cite{1957_TransFaradaySoc_Bradley, 1958_ProcRSocLondA_Eigen, 1984_ApplPhysA_Langer, 2013_SciRep_Deng}, in particular in the state of ice where the oxygen centers are comparatively immobile. Pure water corresponds to an intrinsic semiconductor with protonic valence and conduction bands that are separated by a protonic band gap~\cite{1984_ApplPhysA_Langer}. Thermal excitation across the band gap generates the intrinsic $10^{-7}\,\mathrm{M}$ concentration of mobile protons ($\mathrm{H^{+}}$) and proton holes, i.e. $\mathrm{OH^{-}}$. Acids and bases correspond to proton donor and acceptor dopants that shift the protonic Fermi level closer to the conduction and valence bands, respectively~\cite{1984_ApplPhysA_Langer}. This perspective on aqueous acid--base electrolytes matches the MCDFT formalism, where the protonic band structure is reflected in the protonic DOS of the Kohn-Sham orbitals. The present results suggest that the aqueous protonic DOS, or, more generally, the ionic DOS, can be probed by measuring the capacitance of systems with geometrically confined electrolyte phases, such as electrolyte pores~\cite{2013_FaradayDiscuss_Kornyshev, 2020_JChemPhys_Eikerling}, channels~\cite{2018_Science_Fumagalli}, or thin-films~\cite{2019_NatComm_Yamada, 2021_NatMater_Boyd}. For macroscopic electrode and electrolyte domains, on the contrary, the DOS capacitance contribution to the total interface capacitance in equation~\eqref{eq_total_capacitance_MCDFT} becomes negligible, since the electronic and ionic DOS scale with the size of the electrode and electrolyte regions, respectively.

\subsection{Intercalation materials and pseudocapacitance}
\label{subsec_discussion_intercalation}

So far, the discussion was focused on electrochemical interfaces. However, the present formalism does not necessarily require electrode and electrolyte to represent spatially separate regions. It also applies to mixed electron-ion conductors and active battery materials, where mobile electrons and ions are present within the same phase. In such a case, the inverse capacitance corresponds to the voltage slope~\cite{2016_ChemMater_Ceder} of the battery (dis)charging profile, see equation~\eqref{eq_capacitance_MCDFT}. For a Li-intercalation material, e.g., the electrons and Li-ions are conveniently treated within the MCDFT framework, whereas the other atom cores of the crystal lattice represent the external charge density. We therefore only have one ionic species, i.e. $\mathrm{Li^{+}}$, and $t_{\mathrm{Li^{+}}} = 1$ according to equation~\eqref{eq_transference_coefficients_sum}. The reference electrode reaction is $\mathrm{Li} - \mathrm{Li^{+}} \rightleftharpoons \mathrm{e^{-}}$, so $\alpha_{\mathrm{Li^{+}}} = -1$. The electrostatic capacitance of equation~\eqref{eq_el_capacitance_MCDFT} becomes
\begin{eqnarray}
\label{eq_el_capacitance_battery}
\frac{e^2}{C_{\mathrm{el}}} = \int\left[ -e\,\Gamma^T_{\mathrm{LDOS}}(\mu^{\mathrm{eff}},\mathbf{r}) + e\,\Gamma^T_{\mathrm{LDOS},\mathrm{Li^{+}}}(\mu_{\mathrm{Li^{+}}}^{\mathrm{eff}},\mathbf{r})\right]\frac{\partial \phi(\mathbf{r})}{\partial N}\mathrm{d}\mathbf{r} 
\end{eqnarray}
Unlike for the case of an electrochemical interface, the potential response $\partial \phi(\mathbf{r})/\partial N$ does \emph{not} assume a constant plateau value in the bulk of an active electrode material. It follows from equation~\eqref{eq_electrostatic_potential_MCDFT}:
\begin{eqnarray}
\label{eq_electrostatic_potential_response}
\frac{\partial \phi(\mathbf{r})}{\partial N} = \frac{1}{4\pi\epsilon_0}\,\int \frac{1}{|\mathbf{r}-\mathbf{r}'|}\left[-e\,\frac{\partial n(\mathbf{r}')}{\partial N}+ e\,\frac{\partial n_{\mathrm{Li^{+}}}(\mathbf{r}')}{\partial N}\right]\,\mathrm{d}\mathbf{r}'
\end{eqnarray}
The electronic and ionic density response $\partial n(\mathbf{r}')/\partial N$ and $\partial n_{\mathrm{Li^{+}}}(\mathbf{r}')/\partial N$, i.e. the added charge, extend throughout the bulk of the active material. Inserting equation~\eqref{eq_electrostatic_potential_response} into \eqref{eq_el_capacitance_battery}, we see that the inverse of the electrostatic capacitance is equal to the electrostatic interaction between the density response and the local density of states, both including the respective electronic and ionic contributions.

\begin{figure}[t]
\centering
\includegraphics[width=0.8\textwidth]{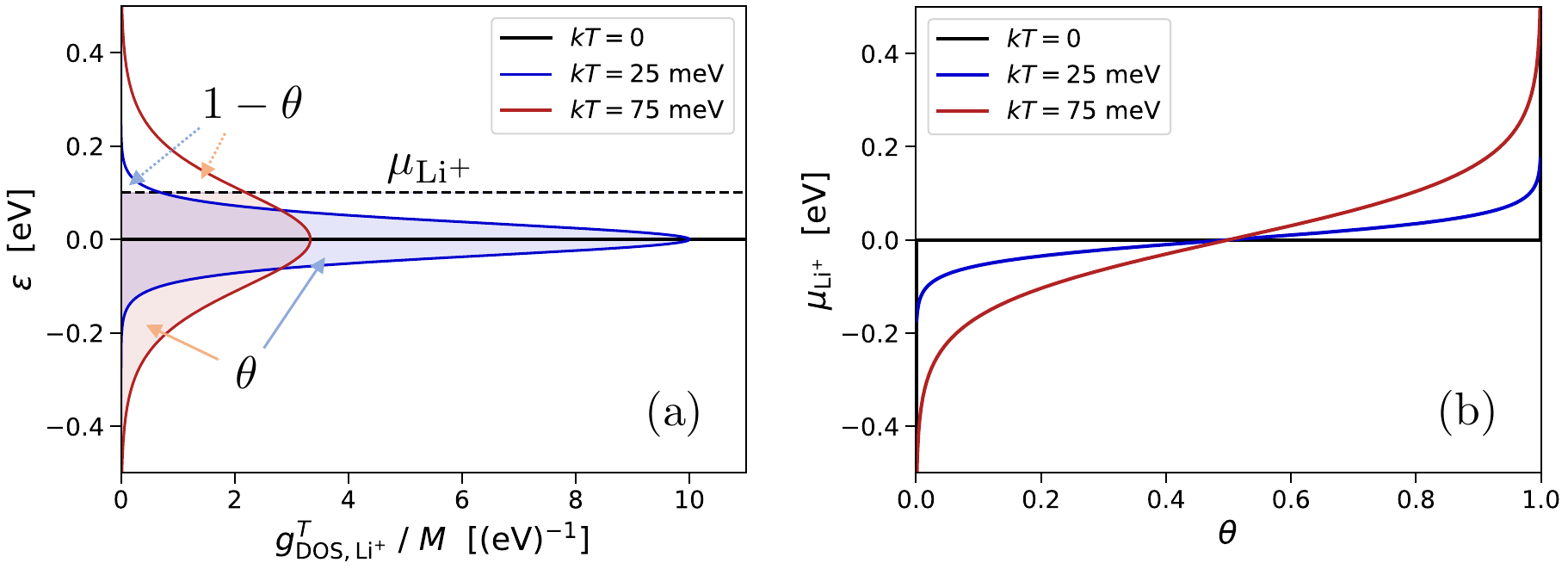}
\caption{Kohn-Sham treatment of Li-ion intercalation: (a) Schematic plot of the temperature-dependent ionic DOS $g^T_{\mathrm{DOS},\mathrm{Li^{+}}}(\epsilon)$ normalized to the number of intercalation sites $M$ at different temperatures. The site (orbital) energy is set to $\epsilon_{\mathrm{s}}=0$. The thermally broadened DOS is filled up to the ionic Fermi energy $\mu_{\mathrm{Li^{+}}}^{\mathrm{eff}}$. (b) Charging profile: Site coverage $\theta$ versus ionic Fermi energy $\mu_{\mathrm{Li^{+}}}^{\mathrm{eff}}$ in the frozen orbital approximation at different temperatures. The curves are identical to the ones obtained from the classical statistical treatment.}
\label{fig_ionic_DOS}
\end{figure}

The DOS capacitance~\eqref{eq_DOS_capacitance_MCDFT} of a Li-ion intercalation material reads
\begin{eqnarray}
\label{eq_DOS_capacitance_battery}
\frac{1}{C_{\mathrm{DOS}}} \,=\, \frac{1}{C_{\mathrm{DOS},e}} + \frac{1}{C_{\mathrm{DOS},\mathrm{Li^{+}}}}
\end{eqnarray}
Here, the ionic DOS capacitance $C_{\mathrm{DOS},\mathrm{Li^{+}}}$ includes the \emph{configurational} density of states of the partially occupied Li-ion sites within the host crystal lattice. It can be readily shown that the present formalism is consistent with the classical statistical treatment. For this, we consider a host lattice with a number $M$ of symmetrically equivalent sites, occupied by a total of $N$ $\mathrm{Li^{+}}$ ions. The occupation of any site corresponds to a change in internal energy by the site energy $\epsilon_{\mathrm{s}}$. At the simplest level, the interaction between occupied sites is neglected and $\epsilon_{\mathrm{s}}$ is constant. In the classical treatment, the free energy of the system is then written as $A(N) = N\epsilon_{\mathrm{s}} - kT \log\binom{M}{N}$, where the second term results from the configurational entropy of distributing the $N$ ions over the $M$ available sites. Using Stirling's formula, one immediately obtains the well-known relation for the chemical potential $\mu = \partial A/\partial N = \epsilon_{\mathrm{s}} + kT\log\left(\frac{\theta}{1-\theta}\right)$, where $\theta = N/M$ is the site coverage. The capacitance then follows as $C = e^2(\partial \mu/\partial N)^{-1} = M(e^2/kT)\,\theta\,(1-\theta)$. The same result is obtained from the present formalism. Here, each intercalation site corresponds to one ionic Kohn-Sham orbital. There are $M$ symmetrically equivalent orbitals and, accordingly, $M$ degenerate energy eigenvalues $\epsilon^{\mathrm{Li^{+}}}_{\kappa} = \epsilon_{\mathrm{s}}$ for $\kappa = 1,\dots,M$. The neglect of site--site interactions corresponds to the frozen orbital approximation, i.e. $\partial\epsilon^{\mathrm{Li^{+}}}_{\kappa}/\partial N = 0$, and the same for the electronic Kohn-Sham eigenvalues. It then directly follows, see equation~\eqref{eq_chem_pot_deriv_I}, that the electrostatic and XC terms at the right-hand side of equations~\eqref{eq_chem_pot_deriv_electron_MCDFT} and \eqref{eq_chem_pot_deriv_ion_MCDFT} are zero and the total system capacitance is given by the DOS capacitance, $C = C_{\mathrm{DOS}}$. Furthermore neglecting the electronic contribution, i.e. assuming a sufficiently large value of the electronic DOS, we have $C \approx C_{\mathrm{DOS},\mathrm{Li^{+}}}$, see equation~\eqref{eq_DOS_capacitance_battery}. Because of the degeneracy of eigenvalues, condition~\eqref{eq_KS_ion_number_MCDFT} for the Fermi-Dirac weights reads $N = M/\left[1+\exp\left((\epsilon_{\mathrm{s}}-\mu_{\mathrm{Li^{+}}}^{\mathrm{eff}})/kT\right)\right]$, where we also used $N_{\mathrm{Li^{+}}} = N$ due to charge neutrality. Resolving for the chemical potential, we obtain the same expression as from the classical treatment, $\mu_{\mathrm{Li^{+}}}^{\mathrm{eff}} = \epsilon_{\mathrm{s}} + kT\log\left(\frac{\theta}{1-\theta}\right)$, shown in Figure~\ref{fig_ionic_DOS}(b) at different temperatures. The ionic DOS reads $g^T_{\mathrm{DOS},\mathrm{Li^{+}}}(\epsilon) = M\,p^T(\epsilon-\epsilon_{\mathrm{s}}) = (M/kT)\,\exp[(\epsilon-\epsilon_{\mathrm{s}})/kT]/\left(1+\exp[(\epsilon-\epsilon_{\mathrm{s}})/kT]\right)^2$, which is shown in Figure~\ref{fig_ionic_DOS}(a) at different temperatures. When evaluated at $\epsilon = \mu_{\mathrm{Li^{+}}}^{\mathrm{eff}}$, it yields the ionic DOS capacitance $C_{\mathrm{DOS},\mathrm{Li^{+}}} = e^2\,g^T_{\mathrm{DOS},\mathrm{Li^{+}}}(\mu_{\mathrm{Li^{+}}}^{\mathrm{eff}}) = M(e^2/kT)\,\theta\,(1-\theta)$. The Fermi-Dirac distribution on the ionic Kohn-Sham eigenvalues thus leads to the same result as the classical statistical treatment of Li-ion intercalation. The same then also holds for the description of electrochemical adsorption to a discrete number of sites at the surface of an electrode material. This demonstrates that the MCDFT formalism provides a unified theoretical framework for the first-principles description of electrochemical capacitance, including double-layer charging as well as the pseudocapacitance due to electrochemical adsorption processes and the bulk capacitance of active battery materials.

\begin{figure}[t]
\centering
\includegraphics[width=1.0\textwidth]{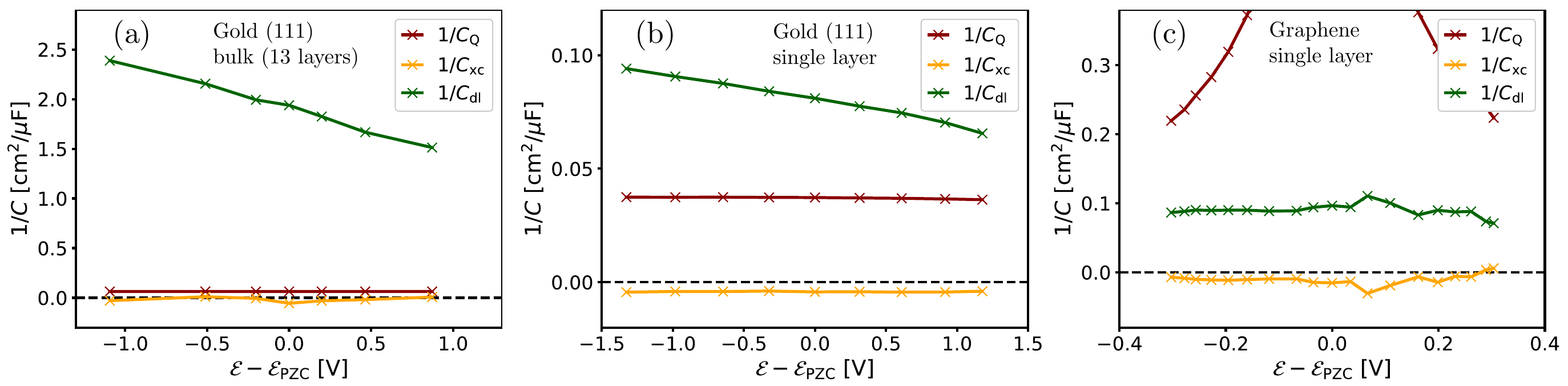}
\caption{Capacitance partitioning: Contributions to the total capacitance according to equation~\eqref{eq_total_capacitance_JDFT} computed at the JDFT level for a 13-layer Au (111) slab electrode (a), single-layer Au (111) electrode charged from both sides (b), and single-layer graphene electrode charged from both sides (c), plotted as a function of potential vs. the respective potential of zero charge ($\mathcal{E}_{\text{PZC}}$). (J)DFT calculations were performed with VASP~\cite{1996_CompMaterSci_Kresse} with PAW pseudopotentials~\cite{1999_PhysRevB_Kresse} and the PBE XC-functional~\cite{1996_PhysRevLett_PBE}, Fermi-Dirac smearing with $kT=50\,\mathrm{meV}$, energy cutoff of $450\,\mathrm{eV}$ for Au and $520\,\mathrm{eV}$ for graphene, and $\Gamma$-centered $K$-point meshes: $(27\times 27\times 1)$ for 13-layer Au (111), $(97\times 97\times 1)$ for single-layer Au (111), $(193\times 193\times 1)$ for graphene. The VASPsol implicit electrolyte model~\cite{2019_JChemPhys_Mathew_Hennig} was used with dielectric constant $\epsilon_{\mathrm{r}}=78.4$ and Debye length $\lambda_{\text{D}}=1.5\,\mathrm{\AA}$. The fixed atomic positions were derived from the relaxed Au bulk lattice and relaxed single-layer graphene in vacuum, respectively.}
\label{fig_XC_capacitance}
\end{figure}

\subsection{Exchange and correlation capacitance}

The existence of the universal exchange and correlation functionals within (MC)DFT is proven~\cite{1964_PhysRev_Hohenberg_Kohn, 1965_PhysRev_Mermin, 1965_PhysRev_Kohn_Sham, 1998_PhysRevB_Gidopoulos}, but their precise form is unknown. However, many useful approximation schemes have been developed especially for the electronic XC functional. Although a detailed mathematical analysis of the exchange and correlation capacitance is beyond the scope of the present article, we will highlight some relevant properties. Figure~\ref{fig_XC_capacitance} shows the computed individual capacitance contributions (plotted as inverse capacitance) for three different electrode--electrolyte systems, with a 13-layer (bulk) gold (111) electrode (a), a single-layer gold (111) electrode (b), and a single-layer graphene electrode (c). The calculations were performed with the VASPsol implicit electrolyte model~\cite{2019_JChemPhys_Mathew_Hennig}, which means that the plotted quantum capacitance $C_{\mathrm{Q}}$ only corresponds to the electronic DOS, see equation~\eqref{eq_quantum_capacitance_JDFT}. Also the plotted XC capacitance $C_{\mathrm{xc}}$ mainly represents the electronic contribution $C_{\mathrm{xc}}^{ee}$ given by equation \eqref{eq_xc_ee_capacitance_MCDFT}, but it additionally includes the contribution of the electrolyte boundary functional of VASPsol~\cite{2019_JChemPhys_Mathew_Hennig}. For the bulk gold electrode (Figure~\ref{fig_XC_capacitance}a), the contribution of the electronic XC capacitance, $1/C_{\mathrm{xc}}$, is negligible. However, for both the single-layer gold and graphene electrodes we find a slightly negative value for $1/C_{\mathrm{xc}}$. This is surprising, because we typically consider the capacitance as a positive quantity. The negative value of the electronic XC capacitance can be explained with the fact that, by construction, the XC functional is defined as the difference between the full universal functional and the other functionals explicitly included in the Kohn-Sham scheme. In particular, the exact (unknown) XC functional comprises the electrostatic self-interaction correction and electrostatic correlation beyond the mean-field approximation, both of which generally facilitate charge localization. From this perspective, it is not surprising that the contribution of the XC functional results in an \emph{increase} in capacitance, which is achieved by the \emph{negative series capacitor} $1/C_{\mathrm{xc}}$, see equations~\eqref{eq_total_capacitance_JDFT} and \eqref{eq_total_capacitance_MCDFT}. This contrasts with the usual picture where a (positive) series capacitor leads to a decreased total capacitance, whereas an increase in total capacitance is usually achieved by connecting (positive) capacitive contributions in parallel. 

A similar behavior is expected for the ionic exchange-correlation capacitance $C_{\mathrm{xc}}^{ii}$, see equation~\eqref{eq_xc_ii_capacitance_MCDFT}, and the electron--ion correlation capacitance $C_{\mathrm{c}}^{eI}$, see equation~\eqref{eq_xc_ei_capacitance_MCDFT}, the respective functionals largely representing corrections of the electrostatic mean-field energy~\cite{1998_PhysRevB_Gidopoulos}. To which extent these capacitance contributions play a role in the (classical) limit of a bulk electrolyte phase remains to be investigated in the future.

\section{Conclusions}

It was shown that multicomponent density-functional theory (MCDFT) is a suitable formalism for the first-principles description of electrochemical capacitance. It enables a unified treatment of both non-faradaic (double-layer charging) and faradaic (pseudocapacitance) contributions. An exact analytical expression for the total capacitance was derived from the electronic and ionic Kohn-Sham equations. Three general capacitance contributions were obtained: First, the ``quantum'' capacitance of the density of states (DOS) of all active species; second, the electrostatic mean-field capacitance; third, the exchange and correlation capacitance. The DOS capacitance is relevant for systems with confined electrode or electrolyte phases. For the case of an intercalation material, it comprises the contribution of the configurational DOS of the partially occupied sites for the active ion species. The electrostatic capacitance is determined by the local density of states (LDOS) of all species and the respective expression was shown to converge to the classical definition of the double-layer capacitance for the case of an electrochemical interface with bulk electrode and electrolyte domains. Computational results indicate that the exchange and correlation capacitance can have negative values, which can be explained in terms of a correction of the mean-field electrostatic contributions. Since it formally acts in series, a negative value would lead to an increase in total capacitance. The present work advocates the advancement of MCDFT for the first-principles modeling of electrochemical systems.

\section*{Acknowledgements}
This work was supported by the European FET-Open project VIDICAT (Grant Agreement: 829145).

\bibliographystyle{unsrt} 
\bibliography{capacitance_first_principles}

\end{document}